\documentclass[12pt]{article}
\usepackage{amssymb,amsmath}
\include{epsf}

\usepackage{graphicx}
\usepackage[arrow,matrix,curve]{xy}

\usepackage{amsmath}
\usepackage{amssymb}
\usepackage{amscd}
\usepackage{amsfonts}
\usepackage{amsthm}

\newcommand{\e}{{e}}
\newcommand{\ii}{{i}}
\newcommand{\dd}{{d}}

\newcommand{\eqn}[1]{(\ref{#1})}

\def\appendix#1{\addtocounter{section}{1}\setcounter{equation}{0}
\renewcommand{\thesection}{\Alph{section}}
\section*{
\thesection\protect\indent \parbox[t]{11.715cm} {#1}}
\addcontentsline{toc}{section}{Appendix\thesection\ \ \ #1} }
\newcommand{\real}{{\mathbb R}} 

\newcommand{\be}{\begin{equation}}
\newcommand{\ee}{\end{equation}}
\newcommand{\beq}{\begin{equation}}
\newcommand{\eeq}{\end{equation}}
\newcommand{\bea}{\begin{eqnarray}}
\newcommand{\eea}{\end{eqnarray}}
\def\beqa{\begin{eqnarray}}
\def\eeqa{\end{eqnarray}}
\def\nn{\nonumber}
\newcommand{\del}{\partial}

\newcommand{\eq}{\begin{equation}}
\newcommand{\eqa}{\begin{eqnarray}}
\newcommand{\en}{\end{equation}}
\newcommand{\ena}{\end{eqnarray}}

\def\nn{\nonumber}

\newcommand{\al}{\alpha}

\newcommand{\RR}{{\mathcal R}}

\newcommand{\R} {R}

\newcommand{\st}{{\star}}

\newcommand{\FF}{\mathcal F}

\newcommand{\x}{\mathsf{x}}
\newcommand{\y}{\mathsf{y}}

\newcommand{\ff}{{\sf f}}

\newcommand{\rr}{{\R}}

\begin{document}
\begin{titlepage}
\begin{flushright}

\baselineskip=12pt
DSF--19--2008\\
\hfill{ }\\
\end{flushright}

\begin{center}

\baselineskip=24pt

{\Large\bf Noncommutative Conformal Field Theory in the Twist-deformed context}

\baselineskip=14pt

\vspace{1cm}

{\bf Fedele Lizzi$^{2}$ and Patrizia
Vitale$^{2}$}
\\[6mm]
 $^{2}${\it
Dipartimento di Scienze Fisiche, Universit\`{a}
di Napoli {\sl Federico II}}\\ and {\it INFN, Sezione di Napoli}\\
{\it Monte S.~Angelo, Via Cintia, 80126 Napoli, Italy}\\{\small\tt
fedele.lizzi@na.infn.it,
patrizia.vitale@na.infn.it}
\\[10mm]

\end{center}

\vskip 2 cm

\begin{abstract}
We discuss conformal symmetry on the two dimensional
noncommutative plane equipped with Moyal product in the twist
deformed context. We show that the consistent use of the twist
procedure leads to results which are free from ambiguities. This lends support to the importance of the use of
twist symmetries in noncommutative geometry.

\end{abstract}

\end{titlepage}
\section{Introduction}
The presence of a noncommutative structure of space (or spacetime),
can be described by a nontrivial commutation relation among the
coordinates, or more generally by the presence of a deformation of
the product among functions, the deformed $\star$ product. The
presence of the noncommutativity parameter also means that spacetime
symmetries are in general broken, at least as classical symmetries.
It is in fact possible to consider a \emph{quantum symmetry} for
these theories. This means that the Lie Algebra is deformed into a
noncocommutative Hopf algebra, and in this case it is still possible
to have a symmetry, albeit of a new, quantum, type.

While most of the interest has been for the Lorentz or Poincar\'e
symmetries, in this paper we will concentrate our attention on the
two-dimensional conformal symmetry~\cite{BPZ}.

In this paper we will consider a field theory on the noncommutative
plane ${\real}^2_\theta$. This is the plane equipped with a
noncommutative star product for which
\be
x_1\star x_2-x_2\star x_1=\ii\theta
\ee
with $\star$ the Gr\"onewold-Moyal star product
\begin{equation}
(f \star g)(x) = e^{\frac{i}{2}\theta (\partial_{x_0} \partial_{y_1} -
    \partial_{x_1} \partial_{y_0})} f(x) \cdot g(y)|_{x=y}
\end{equation}

In~\cite{LVV06} we showed that, by deforming the commutation
relations between creation and annihilation operators
appropriately, quantum conformal invariance of a simple
two-dimensional field theory is preserved on the noncommutative
plane. In that paper we followed the following philosophy: We
assumed there would be a conformal symmetry, \emph{and then}
looked for commutation relations which could fulfill them. We
found that it was necessary a deformation of the commutator.
Deformations of this kind had appeared earlier in the
literature (for example in~\cite{FioreSchupp1,FioreSchupp2,
bmpv}, and have been not free from controversies. In~\cite{ALV}
together with Paolo Aschieri we have given a consistent
prescription for the application in the twist, which we will
discuss below. In~\cite{paga,nini}, together with S.~Galluccio
we have shown as the consistent use of this procedure enables
the understanding, at a physical level, of the equivalence
among the Gr\"onewold-Moyal and the Wick-Voros product. In this
note we reconsider the issue of twisted conformal symmetry and
show that the consistent use of the procedure developed
in~\cite{ALV} leads to results which are free from ambiguities
and consistent with~\cite{LVV06}.

\section{The model}
\setcounter{equation}{0}

We first present, mainly to fix notation, the model in the
usual commutative setting. Let us consider the two dimensional
Minkowski plane in light cone coordinates.  We will use the
convention in which the spacetime metric $\eta_{\mu \nu} ={\rm
diag}(1, -1)$. Light cone coordinates are defined as $x^\pm =
x^0 \pm x^1$. Using $\eta_{AB} dx^A dx^B = \eta_{\mu \nu}dx^\mu
dx^\nu$, and $\eta^{AB}=(\eta_{AB})^{-1}$ (here $A, B$ label
lightcone indices + and -) we have
\begin{eqnarray}
\eta_{++} = \eta_{--} &=0=& \eta^{++} = \eta^{--}, \nonumber \\
\eta_{+-} = \eta_{-+} &= \frac{1}{2}=&(\eta^{+-})^{-1} =
(\eta^{-+})^{-1}.
\end{eqnarray}
{}From $x_A = \eta_{AB} x^B$ we also have the rule $ x_\pm =
{x^\mp}/{2}\ .$

We consider the simplest two dimensional conformally invariant
theory, a scalar massless field theory described by the action
\begin{equation}
S=\int d^2x ~ \del_+\varphi  \del_-\varphi \ .
\end{equation}
  The classical solutions are fields split in
``left'' and `right'' movers $
\varphi=\varphi_+(x^+)+\varphi_-(x^-)$. In the quantum theory the
fields are operators with mode expansion
\be  \phi(x^0,x^1) = \int_{-\infty}^\infty \frac{dk^1}{4\pi
k_0}\left( a(k) e^{-ik^\mu x_\mu} + a^\dag (k)e^{ik^\mu x_\mu}
\right). \ee Using $k^0=k_0=|k^1|$  this can be rewritten as
\bea
 \phi(x^+,x^-) &=& \int_{-\infty}^0 \frac{dk^1}{4\pi |k^1|}\left(
a(k) e^{- i|k^1| x^+} + a^\dag (k)e^{i|k^1| x^+} \right)
\nonumber\\  &&+ \int_{0}^{\infty} \frac{dk^1}{4\pi
|k^1|}\left( a(k) e^{-ik^1 x^-} + a^\dag (k)e^{ik^1 x^-}\right)
\label{phi}.
\eea
This in turn may be rewritten as
\be  \phi(x^+,x^-) =\int_{-\infty}^\infty  \frac{dk^1}{4\pi |k^1|}\left(
a_-(k) e^{- ik^1 x^+} + a_+(k) e^{- i k^1 x^-}\right)=
\phi_+(x^+)+\phi_-(x^-) \label{phipm}
\ee
 where we have introduced the symbol
$a_\sigma(k), ~\sigma=+,-, ~k\in(-\infty,\infty)$ related to the two
sets of oscillators appearing in \eqn{phi} (left and right movers)
by $a_\sigma(k)=a(\sigma k),~ a_\sigma(-k)=a^\dag(-\sigma k),~
k\in(0,\infty)$ . The commutation relations for the creation and
annihilation operators are
\begin{equation}
[a_\sigma(p),a_{\sigma'}(q)]= 2 p\delta(p+q)\delta_{\sigma\sigma'}.
\label{standcomma}
\end{equation}
Then the quantum currents
\beqa
J^+(x)&=&2 J_-(x)=2 \del^- \phi=i \int_0^{\infty} \frac{dk^1}{2\pi}\left(
a^\dag (k)e^{ik^1
  x^-} - a(k) e^{- ik^1 x^-} \right) \nn\\
  J^-(x)&=& 2 J_+(x)=2 \del^+ \phi=i \int_{-\infty}^0 \frac{dk^1}{2\pi}\left(   a^\dag (k)e^{-ik^1 x^+}
- a(k) e^{ ik^1 x^+}\right)\label{KC}
\eeqa
generate two commuting U(1) Ka\c{c}-Moody algebras with central
extension
\begin{eqnarray}
\left[J^{\pm}(x),J^{\pm}(y)\right] &=&
-\frac{i}{\pi}\del_{\mp}
\delta(x^\mp - y^\mp),\nn \\
\left[J^+(x),J^-(y)\right] &=& 0. \label{KM}
\end{eqnarray}
 Quantum conformal
invariance is proved showing that the components of the quantum
stress-energy tensor
\begin{eqnarray}
\Theta_{\pm\pm} &=& \frac{1}{4}(\Theta_{00}\pm2
\Theta_{01}+\Theta_{11}),\nonumber \\
\Theta_{+-} &=& \frac{1}{4}(\Theta_{00} - \Theta_{11}) = \Theta_{-+}
\end{eqnarray}
generate the conformal algebra. Tracelessness and conservation
($\partial^\mu \Theta_{\mu \nu}=0$) imply $ \Theta_{\pm\mp} =0$ and
$\partial^\pm \Theta_{\pm\pm} = 0$. Hence $\Theta_{\pm\pm} (=
\frac{\Theta^{\mp\mp}}{4})$ is a function of $x^\mp$ only, as in the
standard case. Classically, $\Theta^{++}(x) = J^+(x) J^+(x)$. The
quantum stress-energy tensor is the {\it normal-ordered} product
\be
\Theta^{\pm\pm}(x)= \gamma  :J^\pm(x) J^\pm(x):\ , \label{defLquant}
\ee
where $\gamma$ is a real number which gets fixed in the quantum
theory, and normal ordering is defined as
\bea
:a_\sigma(p)a_\sigma(q):&=&a_\sigma(p)a_\sigma(q)\ \mbox{if}\  p<q\nonumber\\
:a_\sigma(p)a_\sigma(q):&=&a_\sigma(q)a_\sigma(p)\ \mbox{if}\  p\geq
q.\label{nord}
\eea

Creation and annihilation operators obey the commutation rules
\begin{equation}
a_\sigma (p) a_{\sigma'}(q) =  2p \delta(p+q)\delta_{\sigma\sigma'},\label{acom}
\end{equation}
Therefore the existence of Ka\c{c}-Moody quantum current algebras is
a sufficient condition to ensure conformal invariance at the quantum
level. It results
\beqa
[\Theta^{\pm\pm}(x),\Theta^{\pm\pm}(y)] & = &\pm \frac{4 i}{\pi}
\Theta^{\pm \pm}(x)\del_{\mp} \delta(x-y) - \frac{i}{6 \pi^3}
\del_{\mp}^{'''} \delta(x-y)
\nonumber \\
 {[}\Theta^{++}(x),\Theta^{--}(y){]} &&=0 \label{confal}
\eeqa
.
\section{The NC case: a review}
\setcounter{equation}{0}

We consider now Moyal type noncommutativity, which implies, for
the fields of the theory
\begin{equation}
\phi(x)\star \psi(y)= \left[m_0 \circ {\cal F}^{-1} (\phi\otimes
\psi)\right](x,y)=e^{\frac{i}{2}\theta^{\mu
    \nu}\partial_{x_\mu} \partial_{y_\nu}} \phi(x) \psi(y).
\end{equation}
Here  $m_0$ is the ordinary pointwise multiplication while  the
twist operator and its inverse are \eq \label{MWTW}
\FF=\e^{-\frac{\ii}{2}\theta^{\mu\nu}\frac{\partial}{\partial x^\mu}
\otimes\frac{\partial}{\partial x^\nu}}~,~~~
\FF^{-1}=\e^{\frac{\ii}{2}\theta^{\mu\nu}\frac{\partial}{\partial
x^\mu} \otimes\frac{\partial}{\partial x^\nu}}~; \en with
$\frac{\partial}{\partial x^\mu}$ and $\frac{\partial}{\partial
x^\nu}$  globally defined vectorfields on $\real^d$
(infinitesimal translations). Given the Lie algebra $\Xi$ of
vectorfields with the usual Lie bracket
\be
[u,v]:= (u^\mu \partial_\mu v^\nu)
\partial_\nu-(v^\nu\partial_\nu u^\mu)\partial_\mu~,
\ee
and its universal enveloping algebra $U\Xi$, the twist $\FF$ is an
element of $ U\Xi\otimes U\Xi.~$ The elements of $U\Xi$ are sums of
products of vectorfields, with the identification $uv-vu=[u,v]$.

We shall frequently write (sum over $\al$ understood) \eq\label{Fff}
\FF=\ff^\al\otimes\ff_\al~~~,~~~~\FF^{-1}=\bar\ff^\al\otimes\bar\ff_\al~,
\en so that \eq\label{fhfg} f\st g:=\bar\ff^\al(f)\bar\ff_\al(g)~.
\en Explicitly we have \eq
\FF^{-1}=\e^{\frac{\ii}{2}\theta^{\mu\nu}\frac{\partial}{
\partial x^\mu} \otimes\frac{\partial}{\partial x^\nu}} =\sum
\frac{1}{n!}\left( \frac{\ii}
2\right)^n\theta^{\mu_1\nu_1}\ldots\theta^{\mu_n\nu_n}
\partial_{\mu_1}\ldots\partial_{\mu_n}\otimes
{}\partial_{\nu_1}\ldots\partial_{\nu_n}=\bar\ff^\al\otimes\bar\ff_\al~,\label{faexp}
\ee
so that $\al$ is a multi-index.

The strategy we followed in \cite{LVV06} was to {\it replace
commutators with deformed commutators}
\be
[a,b]\longrightarrow a\star b-b\star a \label{defocom}
\ee
 and impose that the
KM algebra \eqn{KM} be the same. In the commutative case this was a
sufficient condition for quantum conformal invariance, thanks to
Sugawara construction.  Imposing that
\begin{eqnarray}
J^{\pm}(x)\star J^{\pm}(y)- J^{\pm}(y)\star J^{\pm}(x)&=&
-\frac{i}{\pi}\del_{\mp}
\delta(x^\mp - y^\mp),\nn \\
J^+(x)\star J^-(y)- J^-(y) \star J^+(x)&=& 0. \label{KMd}
\end{eqnarray}
implies in turn a new commutation rule for the creation and
annihilation operators
\begin{equation}
a_\sigma (p) a_{\sigma'}(q) = \mathcal{F}^{-1}(p,q) a_{\sigma'}(q)
a_\sigma (p) + 2p \delta(p+q)\delta_{\sigma\sigma'},\label{acomm}
\end{equation}
where
\begin{equation}
\mathcal{F}^{-1}(p,q)=e^{-\frac{i}{2} p\wedge q}=e^{- i\theta(|p| q
- |q| p)}\
\end{equation}
Then we could verify, a posteriori, that the quantum conformal
commutation relations \eqn{confal} are still valid in the NC case
that is
\beqa
\Theta^{\pm\pm}(x)\star\Theta^{\pm\pm}(y)-\Theta^{\pm\pm}(y)\star\Theta^{\pm\pm}(x)
& = &\pm \frac{4 i}{\pi} \Theta^{\pm \pm}(x)\del_{\mp} \delta(x-y) -
\frac{i}{6 \pi^3} \del_{\mp}^{'''} \delta(x-y)
\nonumber \\
 \Theta^{++}(x)\star\Theta^{--}(y)-\Theta^{++}(y)\star\Theta^{--}(x)&=&0.
\eeqa
\section{The NC case: a new approach}
\setcounter{equation}{0}

In \cite{LVV06} we pointed out that the choice of the commutator
of the NC theory in the form \eqn{defocom} was not canonical, but
was such that the (twisted) conformal symmetry was recovered.
Indeed other choices are possible, motivated by different
principles, see for example \cite{bmpv,gaewes,helsinki2}.

In \cite{ALV}, together with Paolo Aschieri, we have established a
consistent procedure to deform all bilinear maps appearing in a NC
theory, among them,  commutators. The guiding principle was to
deform  bilinear maps
\be \mu\,: X\times Y\rightarrow Z~~~~~~~~~~~~~~~~~~\ee (where $X,Y,Z$
are vectorspaces, with an action of the twist $\FF$ on $X$ and $Y$),
 by combining them  with the action of the twist. In this way we
obtain a deformed version $\mu_\st$ of the initial bilinear map
$\mu$: \eqa \mu_\st:=\mu\circ
\FF^{-1}~,\label{generalpres}&~~~~~~~~~~~~~~& \ena {\vskip -.8cm}
\eqa
{}~~~~~~~~~~~~~\mu_\st\,:X\times  Y&\rightarrow& Z\nn\\
(\x, \y)\,\, &\mapsto&
\mu_\st(\x,\y)=\mu(\bar\ff^\al(\x),\bar\ff_\al(\y))\nn~. \ena The
$\st$-product on the space of functions is recovered setting
$X=Y={\mathcal A}=\mathrm{Fun}(M)$. Without entering into the
description of the general theory, which we don't need here, let us
concentrate on the implications of \eqn{generalpres} for
commutators.

Quantum observables are operator valued functionals defined on
quantum fields. As previously, we choose to work with
light-cone coordinates. Let us consider for a moment the
quantum currents $J^{\pm}(x)$: their undeformed algebra is
given in \eqn{KM}. As we easily verify, their commutator is not
well defined anymore, because it is  not skew-symmetric, it
does not satisfy the Leibniz rule nor Jacobi identity. The
current algebra is trivially derived by the commutation rules
for the left and right component of the field $\phi$
(cfr.~\cite{Heinzl} although with a different normalization)
\begin{eqnarray}
\left[\phi_{\pm}(x),\phi_{\pm}(y)\right] &=&
\frac{i}{8 \pi} {\rm sgn}
(x^\pm - y^\pm),\nn \\
\left[\phi_+(x),\phi_-(y)\right] &=& 0. \label{phipmcom}
\end{eqnarray}
Analogously to $J^\pm$, $\phi_\pm(x^\pm)$ may be regarded as
coordinate fields in the quantum phase space of the theory,
therefore we choose them as as fundamental objects from now on. Of
course their commutator shows the same pathologies as the current
algebra: it is not skew-symmetric, it does  not satisfy the Leibniz
rule nor Jacobi identity. Following our general prescription
\eqn{generalpres} there is a natural commutator on the algebra of
quantum observables which is compatible with the new mathematical
structures. It is obtained by composing the undeformed one with the
twist, appropriately realised on operators.
 We find it convenient to represent the twist
operator \eqn{MWTW} in light cone coordinates
\be
\FF=\e^{-\frac{\ii}{2}\theta^{-+} (\del_{x^-}\otimes
\del_{x^+}-\del_{x^+}\otimes \del_{x^-})}\label{lctwist}
\ee
where we have used
\be
\theta^{ij} \del_i \otimes\del_j=\theta^{-+} (\del_{x^-}\otimes
\del_{x^+}-\del_{x^+} \otimes\del_{x^-}).
\ee
 The algebra of quantum
observables, $\widehat{\mathsf{A}}$, is an algebra of functionals $
G[\phi_\pm]$ on operator valued fields.

We lift the twist \eqn{lctwist} to $\widehat{\mathsf{A}}$ and then
deform this algebra to $\widehat{\mathsf{A}}_\st$ by implementing
the twist deformation principle (\ref{generalpres}). We denote by
$\hat{\partial}_\pm$ the lift to $\widehat{\mathsf{A}}$ of
$\frac\partial{\partial x^\pm}$; for all $  G\in
\widehat{\mathsf{A}}$,
\be
\hat\del_\pm   G:=- \int\dd x^\pm \,\,\del_\pm
\phi_\pm(x)\frac{\delta G}{\delta  \phi_\pm(x)}  \label{dhat}~;
\ee
here $\del_\pm\phi_\pm(x)\frac{\delta   G}{\delta  \phi_\pm(x)}$
stands for $\int\dd\ell\,\del_\pm {\phi_\pm}_{\ell}(x)\frac{\delta
G}{\delta {\phi_\pm}_\ell(x)}$, where like in \eqn{phipm} we have
expanded the operator $\phi_\pm$ as $\int\dd\ell
\,{\phi_\pm}_\ell(x) {\sf a}(\ell)$.

Consequently the twist on operator valued functionals reads
\beqa
\hat{\mathcal F}&=&\e^{-\frac{\ii}{2}\theta^{-+}\left[\!\int\!\dd\!
x^- \left(\del_-  \phi_-\frac{\delta}{\delta
\phi_-(x)}\right)\,\otimes\, \int\!\dd\! y^+\left(\del_+
\phi_+\frac{\delta}{\delta
 \phi_+(y)}\right)-\!\int\!\dd\!
x^+ \left(\del_+  \phi_+\frac{\delta}{\delta
\phi_+(x)}\right)\,\otimes\, \int\!\dd\! y^-\left(\del_-
\phi_-\frac{\delta}{\delta
 \phi_-(y)}\right)\right]}\nn\\
 &\equiv& \bar f^\al \otimes \bar f_\al
\label{lifttwist2}~
\eeqa
and $\bar f$ denote the lift of $\bar \ff$ to operator valued
functionals. This is composed of differential operators of any order
expressed in terms of $\hat \del_\pm$. As we said, in
$\widehat{\mathsf{A}}_\st$ there is a natural notion of
$\st$-commutator, according to the general prescription
\eqn{generalpres}
\be\label{bobobo}
[~,~]_\st= [~,~]\circ \hat {\mathcal F}^{-1}
\ee
that is
\be[F,G]_\st= [\bar f^\al(F), \bar f_\al(G)]~. \label{funcom}
\ee
This $\st$-commutator is $\st$-antisymmetric, is a $\st$-derivation
in ${\hat{\mathsf{A}}}_\st$ and satisfies the $\st$-Jacobi identity
\eqa [  F,  G]_{\st} &= &-[\bar\rr^\alpha(  G),
\bar\rr_\alpha(  F)]_{\st} \label{stantisFc}\\
{}[  F,  G\st   H]_\st &=& [  F,  G]_\st \st   H + \bar\rr^\alpha(
G)\st [\bar\rr_\alpha(  F),  H]_\star
\label{leibtwistFc}\\
{}[  F,[  G,   H]_\star]_\star &=&  [[  F,  G]_\star,  H]_\star +
[\bar\rr^\alpha(  G), [\bar\rr_\alpha(  F),  H]_\star]_\star
\label{jacobitwistFc} \ena Finally, recalling the definition of the
$\RR$-matrix it can be easily verified that
\be\label{stconast}
[  F,  G]_\st=   F\st   G - \bar R^\alpha(  G)\st \bar R_\alpha( F)
\ee
which is indeed the $\st$-commutator in  $\hat {\mathsf{A}}_\st$.

Evaluating the twisted  commutator among the fields we find
\beqa
[\phi_{\pm}(x), \phi_{\pm}(y)]_\st &=&[\phi_{\pm}(x),
\phi_{\pm}(y)]\nonumber\\&& + \frac{i}{2} \theta^{-+} (
[\del_-\phi_{\pm}(x), \del_+\phi_{\pm}(y)]- [\del_+\phi_{\pm}(x),
\del_-\phi_{\pm}(y)])+
O(\theta^2)\nn\\
&=& [\phi_{\pm}(x), \phi_{\pm}(y)] \\
{[\phi_{\pm}(x), \phi_{\mp}(y)]}_{\st}&=&0
\eeqa
because $\phi_\pm$ are respectively functions of $x^\pm$ only. Thus,
the twisted commutators are equal to the undeformed ones. We can
compute the twisted commutators of the currents as well and we find
that those too are undeformed, essentially for the same reason
\beqa
[J^{\pm}(x), J^{\pm}(y)]_\st &=&[J^{\pm}(x),
J^{\pm}(y)]\nonumber\\&& + \frac{i}{2} \theta^{-+} ([\hat-
J^{\pm}(x), \hat\del_+ J^{\pm}(y)]\del_-[\hat\del_+ J^{\pm}(x),
\hat\del_- J^{\pm}(y)]) +
O(\theta^2)\nn\\
&=& [J^{\pm}(x), J^{\pm}(y)] \\
{[J^{\pm}(x), J^{\mp}(y)]}_{\st}&=&0
\eeqa
This  doesn't mean however that $\st$-commutators of more
complicated functionals are undeformed as well. Since the (usual)
Leibniz rule doesn't hold anymore, when we put together functions
their commutator will be different.

To verify, or disprove, quantum conformal invariance, we have to calculate
the twisted commutators  for the energy-momentum tensor components
\be
\Theta^{\pm\pm}(x)= \gamma  :J^\pm(x)\star J^\pm(x):\ ,
\label{thetatwist}
\ee
where the normal ordering is defined in \eqn{nord}.  In order to evaluate these commutators we need the commutation
rules for the creation and annihilation operators. For fixed values of $k$, $a_\sigma(k)$ may be regarded as
functionals of the coordinate fields
\beqa \label{aphi}
a_-(k)&=&2|k^1| \int_{-\infty}^\infty dx^+ e^{ik^1 x^+} \phi_+(x^+) \nn\\
a_+(k)&=&2|k^1| \int_{-\infty}^\infty dx^- e^{ik^1 x^-} \phi_-(x^-)
\eeqa
By means of \eqn{dhat} we find
\be
\hat \del_{\mp}a_{\pm}= -i k^1 a_{\pm}~~~~~~ \hat
\del_{\pm}a_{\pm}=0 \label{dela}
\ee
 Therefore we can compute the
$\st$-commutator of creation and annihilation operators using
\eqn{funcom} and the appropriate expression of the twist
\eqn{lifttwist2}. We find
\beqa
[a_\pm(k), a_{\pm}(k')]_\st &=& [a_\pm(k), a_{\pm}(k')]+
\frac{\ii}{2}\theta^{-+}([\hat \del_- a_\pm(k),\hat \del_+
a_\pm(k')]\nn\\
&-& [\hat \del_+ a_\pm(k),\hat \del_- a_\pm(k')]) + O(\theta^2) =
[a_\pm(k), a_{\pm}(k')]
\eeqa
because of \eqn{dela}, while
\beqa
[a_\pm(k), a_{\mp}(k')]_\st&=&[a_\pm(k), a_{\mp}(k')]+
\frac{\ii}{2}\theta^{-+}([\hat \del_- a_\pm(k),\hat \del_+
a_\mp(k')]\\
&-&[\hat \del_+ a_\pm(k),\hat \del_- a_\pm(k')]) + O(\theta^2)=
e^{-\frac{i}{2}\theta^{-+} k^1 k'^1}[a_\pm(k), a_{\mp}(k')]=0 \nn
\eeqa
because $[a_\pm(k), a_{\mp}(k')]=0$. Therefore the algebra of
creation and annihilation operators is undeformed. Let us consider
now the commutators of the stress-energy tensor components. We have
\beqa
[\Theta^{++}(x),\Theta^{++}(y)]_\st &=&
[\Theta^{++}(x),\Theta^{++}(y)] + \frac{\ii}{2}\theta^{-+}\left(
[\hat\del_- \Theta^{++}(x),\hat\del_+\Theta^{++}(y)] \right.\nn\\
&-& \left.[\hat\del_+
\Theta^{++}(x),\hat\del_-\Theta^{++}(y)]\right)
+ O(\theta^2) \nn\\
&=&[\Theta^{++}(x),\Theta^{++}(y)]
\eeqa
only the zeroth order in $\theta$ survives, that is the undeformed
commutator,  because $\hat\del_\pm\Theta^{\pm}(y)$ is zero. The same
happens for the commutator of the $\Theta^{--}$ component
\beqa
[\Theta^{--}(x),\Theta^{--}(y)]_\st &=&
[\Theta^{--}(x),\Theta^{--}(y)] + \frac{\ii}{2}\theta^{-+}\left(
[\hat\del_- \Theta^{--}(x),\hat\del_+\Theta^{--}(y)] \right.\nn\\
&-&\left.[\hat\del_+ \Theta^{--}(x),\hat\del_-\Theta^{--}(y)]\right)+ O(\theta^2) \nn\\
&=&[\Theta^{--}(x),\Theta^{--}(y)].
\eeqa
Both the
results had to be expected because the twist operator mixes left and
right sectors; therefore surprises could come only from the mixed
commutator
\be
[\Theta^{++}(x),\Theta^{--}(y)]_\st =[\Theta^{++}(x),\Theta^{--}(y)]
+ \frac{\ii}{2}\theta^{-+} [\hat\del_-
\Theta^{++}(x),\hat\del_+\Theta^{--}(y)] + O(\theta^2).
\ee
Let us  compute the first order in the deformation parameter. We
find
\be
\frac{\ii}{2}\theta^{-+} [\hat\del_-
\Theta^{++}(x),\hat\del_+\Theta^{--}(y)]=\frac{\ii}{2}\theta^{-+}\frac{\del}{\del
x^-} \frac{\del}{\del y^+} [\Theta^{++}(x),\Theta^{--}(y)]=0
\ee
because $[\Theta^{++}(x),\Theta^{--}(y)]=0$. The same holds
true for all higher orders in $\theta$, that is the mixed
commutator is undeformed as well and we conclude that, thanks
to the twist, the quantum conformal algebra is preserved.

\end{document}